\documentclass[sigconf,nonacm]{acmart}

\AtBeginDocument{%
  \providecommand\BibTeX{{%
    \normalfont B\kern-0.5em{\scshape i\kern-0.25em b}\kern-0.8em\TeX}}}

\usepackage{transparent}
\usepackage{booktabs}
\usepackage{multirow}
\usepackage{tikz}
\usepackage{array}
\usetikzlibrary{arrows,calc,positioning}

\tikzstyle{intt}=[draw,text centered,minimum size=6em,text width=5.25cm,text height=0.34cm]
\tikzstyle{intl}=[draw,text centered,minimum size=2em,text width=2.75cm,text height=0.34cm]
\tikzstyle{int}=[draw,minimum size=2.5em,text centered,text width=3.5cm]
\tikzstyle{intg}=[draw,minimum size=3em,text centered,text width=6.cm]
\tikzstyle{sum}=[draw,shape=circle,inner sep=2pt,text centered,node distance=3.5cm]
\tikzstyle{summ}=[drawshape=circle,inner sep=4pt,text centered,node distance=3.cm]

\usepackage{float}
\usepackage{soul}
\usepackage{color}
\usepackage{bbm}
\usepackage{enumitem}

\definecolor{Blue}{rgb}{0,0,1}

\definecolor{Orange}{rgb}{1,0.5,0}

\definecolor{Green}{rgb}{0,1,0}

\usepackage[justification=centering]{caption}

\begin{document}

\author{Shachaf Poran}
\email{shachaf.poran@booking.com}
\affiliation{\institution{Booking.com, Tel Aviv}}

\author{Gil Amsalem}
\email{gil.amsalem@booking.com}
\affiliation{\institution{Booking.com, Tel Aviv}}

\author{Amit Beka}
\email{amit.beka@booking.com}
\affiliation{\institution{Booking.com, Tel Aviv}}

\author{Dmitri Goldenberg}
\email{dima.goldenberg@booking.com}
\affiliation{\institution{Booking.com, Tel Aviv}}

\begin{abstract}
Voice assistants provide users a new way of interacting with digital products, allowing them to retrieve information and complete tasks with an increased sense of control and flexibility.
Such products are comprised of several machine learning models, like Speech-to-Text transcription, Named Entity Recognition and Resolution, and Text Classification.
Building a voice assistant from scratch takes the prolonged efforts of several teams constructing numerous models and orchestrating between components. 
Alternatives such as using third-party vendors or re-purposing existing models may be considered to shorten time-to-market and development costs. However, each option has its benefits and drawbacks.
We present key insights from building a voice search assistant for Booking.com search and recommendation system.
Our paper compares the achieved performance and development efforts in dedicated tailor-made solutions against existing re-purposed models.
We share and discuss our data-driven decisions about implementation trade-offs and their estimated outcomes in hindsight, showing that a fully functional machine learning product can be built from existing models.

\end{abstract}


\title[With One Voice: Composing a Travel Voice Assistant from Re-purposed Models]{With One Voice: Composing a Travel Voice Assistant from Re-purposed Models}

\maketitle

\section{Introduction}
\label{sec:intro}

Voice assistants have become a prevailing mode of communication between customers and companies \cite{hoy2018alexa,mari2019voice}. Today you can pick up your smart device and utter a request or a command and the device complies, a thing we wouldn't have dreamt of in the past. The most appealing aspect of this feature is the transfer of touch and typing interfaces into spoken commands, conveying your request in free language and making the action easy to perform and almost instantaneous. For example, you can simply ask a question rather than navigating a verbose FAQ page, or you can use the voice interface when you have limited hand dexterity \cite{whatcanisay}.
Using voice assistants in search and recommendation tasks serves various customer expectations and needs \cite{booking2021personalization}. Introducing a free-form speech input allows customers to generate unstructured queries, resulting in a complex input to the search and recommendation systems \cite{kang2017understanding}.
The unstructured form of natural language also allows users to explore different options in their apps that otherwise would be hidden for the sake of simplicity of the graphical user interface. The user would have to reach these options using buttons and menus that involve more attention and more steps to progress through \cite{guy2016searching}.

A voice assistant relies on a function $v: U\rightarrow A$ that maps an utterance $u \in U$ provided by the user to an action $a\in A$ which can be performed by the app, aiming to fulfill the user's intent which was presented in the utterance. \footnote{Conversational assistants may have additional context which is out scope for this paper.} An example for such a mapping in the groceries domain might be:
\begin{equation*}
     v(\mathit{we\:are\:out\:of\:milk})=\mathit{\:place\:order\:for\:milk}
\end{equation*}
and in the travel domain (which is more relevant in our case) it might be:
\begin{equation*}
    v(\mathit{I\:need\:to\:book\:a\:hotel\:in\:Paris})= \mathit{present\:a\:list\:of\:hotels\:in\:Paris}
\end{equation*}

The actions taken by the Booking.com app may include searching for accommodation, seeking travel inspiration, asking for help, amending existing bookings, recommending a property, etc.

The function $v$ may be seen as a chain of auxiliary functions starting with transforming the raw voice input to text. Only then is the natural language processed to extract the intent of the user and the entities mentioned in the text \cite{8906584,wang2011semantic}. Eventually, a decision is made about which action to perform. In practice, the former two steps are realized using machine learning models.

\begin{figure}[H]
    \includegraphics[width=\columnwidth]{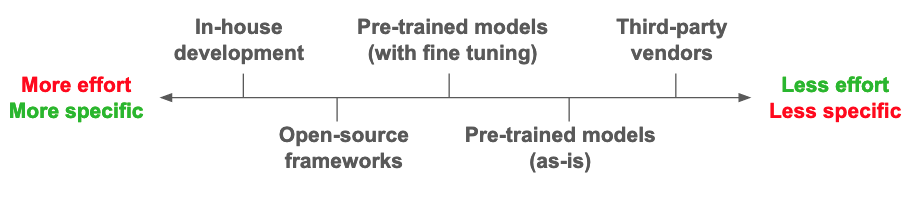}
        \caption{Different options to implementing machine learning models and the trade-off between effort of implementation and specificity of the resulting solution.}
    \label{fig:devtypes}
    \centering
\end{figure}

In creating these machine learning elements, there's a point of decision about how the research and development teams implement them \cite{schalkwyk2010your}. Options include but are not limited to those shown in \autoref{fig:devtypes}.

\begin{figure*}[t]
\centering
  \includegraphics[width=0.95\textwidth]{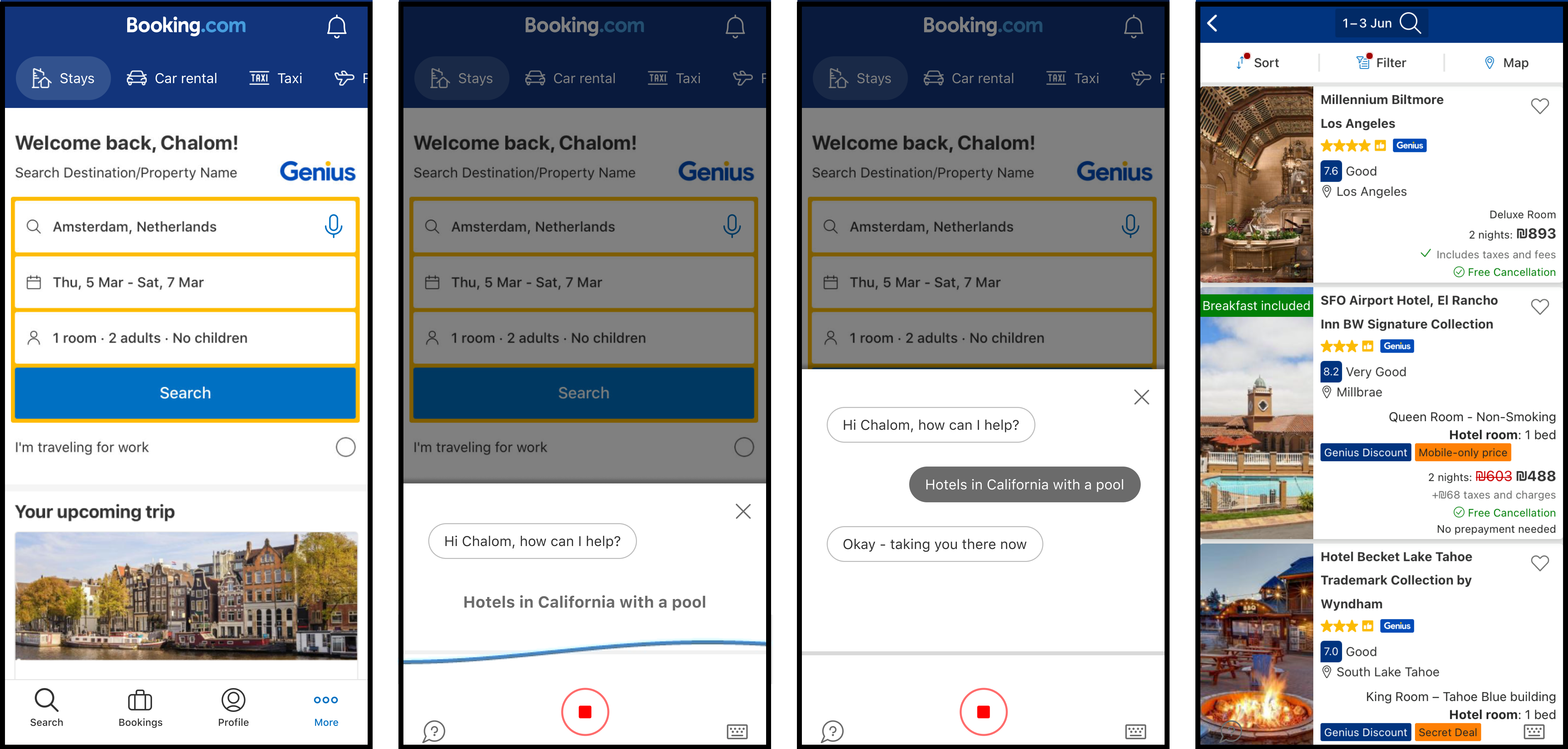}
  \caption{Voice assistant flow, screens from left to right: Index entry point, Search prompt, Input Query, and Search Results.}
  \label{fig:teaser}
\end{figure*}

Each of these options entails implicit costs, whether monetary, development time, or how well the results fit the business needs. 
The items on the left are the ones that are more lengthy and costly in development time, but on the other hand they should also result in more specialized models \cite{schalkwyk2010your}. 
These costs are difficult to estimate in advance and might vary widely depending on the kind of problem to be solved, existing expertise in the workforce, and demand for high accuracy metrics for the models.
Moreover, recent work in the online travel domain has shown that an improvement in an offline metric does not necessarily reflect business impact, and requires online validation via a randomized controlled experiment \cite{haldar2019applying,bernardi2019150}.
At the same time, orchestrating a cascade of machine learning models requires a supporting software system designed to allow a combination of business logic with ML-driven decisions \cite{sculley2015hidden}. 

Another concern when choosing one of these options over another revolves around domain-specific data and knowledge \cite{guy2016searching}. The three leftmost options in \autoref{fig:devtypes} require having data available for training and evaluation, while the other two do not. 
Having the same distribution of data when training a model and when using it for inference is considered good practice, and a significant mismatch between the two might lead to accuracy metrics being irrelevant. Knowledge of these distributions in advance might in some cases lead to using different modeling techniques and better performance.

Constructing a voice assistant usually require a complex architecture, and a generous investment in research and development \cite{kepuska2018next, he2019streaming}. At the same time, re-purposing existing ML models towards new applications \cite{pan2009survey} becomes a popular solution for various product needs. We suggest to adopt a well-known software reuse paradigm \cite{fuqing1999software}, that allows to achieve high quality and reduce development time \cite{li2007empirical} by re-purposing existing machine learning components or considering using external third-party off-the-shelf services \cite{borg2019selecting,petersen2017choosing}.
In this paper we share insights regarding these challenges and how decisions were made in the process of developing a mobile voice assistant (see \autoref{fig:teaser} for an overview of the product flow).
\footnote{app is available at \url{https://www.booking.com/apps.en-gb.html}}

Our key contributions are evidence-based comparisons of dedicated tailor-made solutions against re-purposing existing models for different machine learning tasks. The paper demonstrates how to overcome the lack of in-domain data and to compose a machine learning product without training new models, all the while not compromising potential impact. We share and discuss our data-driven decisions about implementation trade-offs and their estimated outcomes in hindsight by examining the main four components of the system.

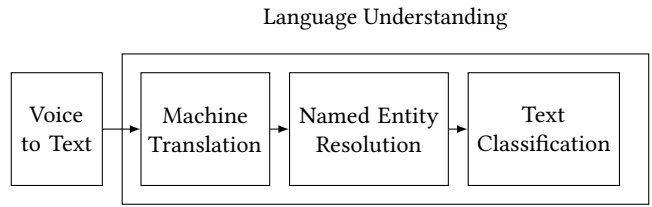
\begin{figure}[b]
    \centering
    \begin{tikzpicture}[
      >=latex,
      node distance=.25cm,
      every node/.style={rectangle,draw,align=center,minimum height=1.5cm},
      auto
    ]
        \node [text width=1cm] (vtt) {Voice to Text};
        \node [text width=1.5cm,right=0.5cm of vtt] (mt) {Machine Translation};
        \node [text width=1.9cm,right=of mt] (ner) {Named Entity Resolution};
        \node [text width=1.8cm,right=of ner] (tc) {Text Classification};
        \node [right=of vtt,minimum width=7cm,minimum height=2cm,label={[label distance=-3mm]above:{Language Understanding}}] (nlp) {};
        
        \draw[->] (mt) -> (ner);
        \draw[->] (ner) -> (tc);
        \draw[->] (vtt) -> (mt);
    \end{tikzpicture}
    \caption{Overview of the voice assistant architecture.}
    \label{fig:vapipeline}
\end{figure}

Voice assistant systems are composed of a voice-to-text element followed by a language understanding element \cite{asrnlp}. The ML pipeline we developed is summarized in the flowchart shown in \autoref{fig:vapipeline}. The voice-to-text (VTT) system is discussed in \autoref{subsec:vtt}. For our use case we chose to construct the language understanding element with three steps: Machine Translation (MT), Named Entity Resolution (NER), and Text Classification (TC). These steps are discussed in \autoref{subsec:translation}, \autoref{subs:ner}, and \autoref{subs:tc} respectively. The output of the last element is fed into a downstream recommender systems \cite{mavridis2020beyond}. \autoref{sec:disscu} concludes our findings and discusses opportunities for future research.  

\section{Voice-to-text}
\label{subsec:vtt}

The first ML-driven element of the pipeline has an utterance as a waveform as input and outputs transcribed text for the uttered speech. It is worthwhile discerning how the distribution of inputs may vary between domains, as it may determine the performance of pre-trained models versus models that were trained on in-domain data. For example:

\begin{enumerate}
    \item \textbf{Sampling rate} values may be either 8KHz or 16KHz.
    \item \textbf{Background noises} such as mechanical hum and vehicle noise for car assistants, colleagues chatter for call-centers, etc.
    \item \textbf{Accents} such as differences in American vs British English.
    \item \label{item:dist} \textbf{Word distribution} differs in different contexts.
\end{enumerate}

Item \ref{item:dist} in the list is especially relevant as VTT systems use statistics of words and n-grams to decide their likelihood. Different domains may exhibit differences in word frequencies that affect accuracy. Even when disregarding domains, different dialects may have a similar effect.

We evaluated two options for the VTT element, one is the open-source framework Kaldi \cite{Povey_ASRU2011} which comes out-of-the-box with ready-made models and tools to tweak them, and the other is a third-party vendor (TPV).
Prior comparisons between Kaldi and an off-the-shelf third-party vendor tool \cite{kimura2018comparison} have shown higher accuracy for Kaldi when testing on in-domain data and when the signal-to-noise ratio is high.

\begin{table}[t]
    \centering
    \begin{tabular}{
    >{\raggedleft}m{0.25\columnwidth}
    >{\raggedright}m{0.2\columnwidth}
    |
    >{\centering}m{0.15\columnwidth}
    >{\centering\arraybackslash}m{0.15\columnwidth}
    }
        \toprule
        Data source & Error word & TPV & Kaldi \\
        \midrule
        Conversations & All words & \textbf{25.25\%} & 28.99\% \\
        App commands & All words & 45.24\% & \textbf{38.68\%} \\
        \midrule
        App commands & "booking" & 198/415 (47.7\%) & \textbf{31/415 (7.5\%)} \\
        App commands & "cancellation" & 46/108 (42.6\%) & \textbf{23/108 (21.3\%)} \\
        \bottomrule
    \end{tabular}
    \caption{Comparison of WER for the TPV vs. the Kaldi model on the adjacent domain and in-domain data-sets.}
    \label{tab:wer}
\end{table}

\begin{table}[b]
    \centering
    \begin{tabular}{c|c}
        \toprule
        TPV & Correct \\
        \midrule
        Contact hotel for \emph{registration} details & reservation \\
        Can I have the \emph{information} & confirmation \\
        \emph{Consolation} & cancellation \\
        \bottomrule
    \end{tabular}
    \caption{Examples of domain-specific errors from the TPV which the Kaldi model got correct.}
    \label{tab:commonerrors}
\end{table}

Developing any model without data generated from the end product produces a classical "chicken or the egg" problem since we cannot infer data distribution. A common practice in this scenario is to use data from an adjacent domain or product to train models. We obtained recordings from customer-service conversations for bootstrapping. Using an annotation tool built in-house, we collected ground-truth transcriptions for these conversations and used them to compare the different models. The metric we used was Word Error Rate (WER) \cite{1318504,niessen-etal-2000-evaluation}, defined as the edit distance between ground truth and predicted sentences normalized by the length of ground truth. This is a common metric used in voice-to-text systems. 

Both TPV and Kaldi allow for the tweaking of their models: the former receives a set of hint phrases that may appear in the utterance and the latter allows fine-tuning modular steps in its VTT pipeline including an acoustic model, a lexicon, and a language model. We tweaked both of the alternatives by using our adjacent-domain data to achieve the lowest WER we could with either.
Kaldi's out-of-the-box model achieved $45.01\%\:WER$, compared to $25.25\%\:WER$ by the TPV. The effort to tweak Kaldi model resulted in $28.99\%\:WER$, resembling similar comparisons with open-access datasets \cite{kimura2018comparison}. Tweaking TPV resulted in a negligible boost in performance. At this point, a decision based on currently-available data was made to use the TPV and defer any additional Kaldi development indefinitely.

After releasing the voice assistant feature real-world data was gathered and annotated, and the two models were reevaluated based on it. \autoref{tab:wer} reports both evaluations, showing that the performance is better for the Kaldi model for utterances taken directly from the product. The same table presents error rates for specific words in the text, explaining some of the difference in performance between the two datasets by the higher abundance of these domain-specific words in the latter.
\autoref{tab:commonerrors} shows common errors by TPV that were transcribed accurately by the Kaldi.

\section{Machine Translation}
\label{subsec:translation}

The work described in \autoref{subsec:vtt} focused on English. When expanding to new markets, the voice assistant is expected to support local languages. Every new language once again faces the same problems already discussed in the previous section, and the time and effort to create the relevant models does not scale well as practically all stages should be repeated, including data collection and annotations.
Using the TPV allowed us to transcribe numerous languages easily, but downstream models were all trained using English inputs. Lack of multilingual training data presented a serious hold-back, which led us to translate the transcriptions before passing them forward \cite{heigold2013multilingual}.

An in-house team has been developing in-domain translation models described in \cite{levin2017machine}.
These models showed consistent results independent of sentence length, which hints that using it for our use case is acceptable. We easily interfaced with their service and served multiple languages with nearly zero effort.

The incisive time to enable new languages has proven essential for testing new markets. Aside from model performance which may differ for each language, user habits for using voice assistants vary with country and culture. Presenting the product to users was key to understanding product-market fit \cite{barnard2010voice,bentley2018understanding}.

\section{Named Entity Resolution}
\label{subs:ner}

Named Entity Recognition (NER) is a Natural Language Processing (NLP) task that asks the question about a word or a sequence of words whether they are "a person, a place or a thing". Named Entity \emph{Resolution} is a task that asks "what person, place, or thing is it". In our context, resolution matches a recognized entity to a closed set of destinations such as countries, cities, villages, and hotels.
Any human hearing the utterance "I'm looking for a hotel in Amsterdam" will assume the speaker intends to visit the Dutch city. However, we are aware that there are different Amsterdams in the United States, South Africa, and elsewhere. Furthermore, we expect two entities to be resolved when we fulfill searches for flights, both for origin and destination.

Entity Resolution is a highly specialized task involving annotation of a substantial amount of in-domain data for both recognition and resolution sub-tasks \cite{hoffart2011robust}. This task is essential for a voice assistant in the travel domain. However, anything other than using a ready-made solution would be infeasible and would delay deployment of the product for a long time.

An in-house team has been developing an Entity Resolution API based on the FLAIR framework \cite{akbik2019flair} for use in a chat-bot product. By the time we came to use it for the voice assistant, it was at near-SOTA performance with more than $90\%$ F1 score for the recognition task. We performed a qualitative inspection and interfaced with the API. This has accelerated our time-to-market, allowing us to present the finalized product to our users quickly.

\begin{figure}[t]
    \includegraphics[width=1\columnwidth]{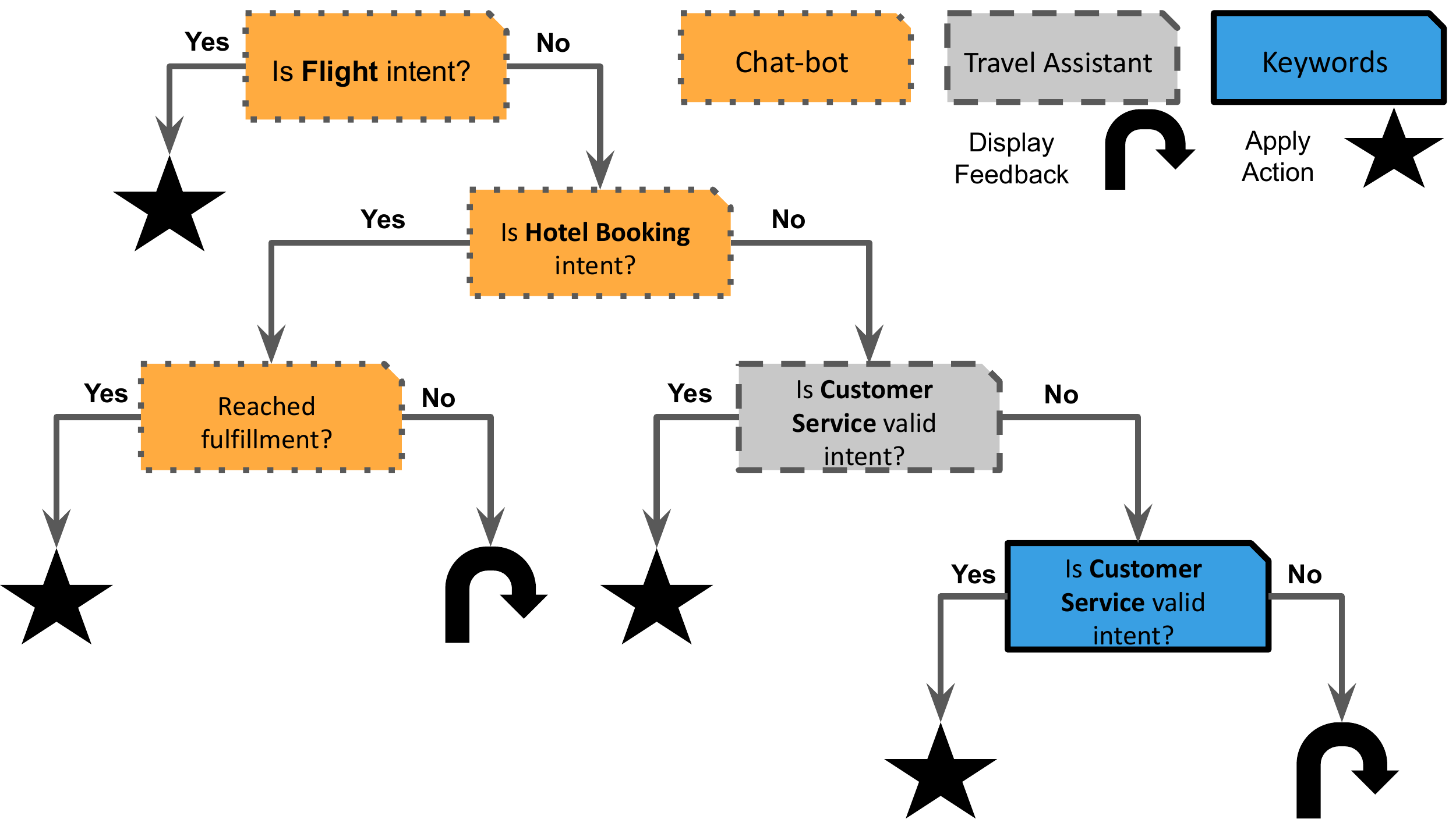}
        \caption{The business logic used to align different classifiers for the text classification task.}
    \label{fig:valogic}
    \centering
\end{figure}

\section{Text Classification}
\label{subs:tc}

\begin{table}[b]
    \centering
    \begin{tabular}{c|c}
        \toprule
        Intents & Prevalence \\
        \midrule
        Pre-book intents        & 66.9\%    \\
        Request human agent     & 8.7\%     \\
        Check booking status   & 7.1\%      \\
        Payments                & 3.0\%     \\
        Change booking          & 1.9\%     \\
        Other post-book intents & 10.0\%     \\
        Greetings               & 2.4\%     \\
        \bottomrule
    \end{tabular}
    \caption{Distribution of intents in our annotated data.}
    \label{tab:topicbreakdown}
\end{table}

In this step of the pipeline, the text is fed into a multi-class classification model and converted into enumerated cases to be used by the client to initiate the response appropriate for the user's utterance. Some of the cases were treated as templates and fulfilled with entities resolved from the NER model, for example searching for an accommodation was fulfilled with the destination.

As a free-form input method, we expected utterances that address both searching for an accommodation to book ("pre-book" intents) and for treating existing bookings ("post-book" intents). User surveys confirmed that, with a distribution of 50\% pre-book intents, 30\% post-book, and the rest are other intents such as greetings and nonsensical queries. This revealed that we have two main sub-domains to address when building the text classification.
Once again, training any model before collecting any data is not feasible. To allow product development and eventually lead to data collection we used two different internal models that serve these features:
\begin{itemize}
    \item \textbf{Travel Assistant}: a text-input interface used to guide users through the FAQ on the site and app. Their NLP model maps text to post-book intents \cite{khalil2019cross}
    \item \textbf{Chat bot}: the tool described in \autoref{subs:ner}. As support to the NER model, it used a different model to decide whether a user wants to book a flight or a hotel (or neither).
\end{itemize}
Interlacing these models using simple rules allowed us to efficiently serve both pre-book and post-book sub-domains with one client-facing interface.
The logic we used for combining the two into a single cohesive product is shown in \autoref{fig:valogic}. Simple if-else statements based on the two models result in either an action such as a flight search being conducted, or an FAQ page being opened, or in giving the user feedback within the voice UI element asking for clarification or additional information. We complemented the process with an exact keyword search, such as \emph{credit} being mapped to \emph{payment} intent, for words we found are significantly correlated with customer-service intents. \emph{coronavirus} is yet another example for such a keyword, which forwarded users to an explanation about the COVID-19 situation in regard to their bookings. Keyword matching works exceptionally well for our use case as the upstream steps filter out most of the other intents.

After the voice assistant feature was made available to the customers, we collected data directly from their interactions and annotated it. Intent distribution, excluding $40\%$ of unintelligible utterances, is given in \autoref{tab:topicbreakdown}. We proceeded to build a model to map text directly to intent using the NLP framework spaCy \cite{spacy}.
The classification metrics to compare the composite business model to the spaCy model are shown in \autoref{tab:topicperf}. These two options were tested in an randomized controlled experiment with two groups, each exposed to a different text classifier. Measuring The number of customer service tasks that were handled by representatives for each of the groups confirmed that the spaCy model results in a reduction in such human-handled tasks which was statistically significant.
\begin{table}[h]
    \centering
    \begin{tabular}{c | c c c c}
        \toprule
        Intent  & \multicolumn{2}{c}{Composite}  & \multicolumn{2}{c}{spaCy} \\
                & p & r                          & p & r \\
        \midrule
        Cancel booking  & \textbf{79\%} & \textbf{70\%} & \textbf{79\%} & 64\% \\
        Change booking  & 79\%          & \textbf{48\%} & \textbf{87\%} & 31\% \\
        Payments        & 46\%          & 50\%          & \textbf{52\%} & \textbf{75\%} \\
        \bottomrule
    \end{tabular}
    \caption{Per-class precision (p) and recall (r) of topic classification models on the most common intents.}
    \label{tab:topicperf}
\end{table}

\section{Conclusion}
\label{sec:disscu}
A common perception of the Data Scientists' work is that their first order of business is training machine learning models to fit the specific tasks at hand, starting with data gathering and ending in a custom model.
Conversely, in the first version of the voice assistant that was released we have not used any machine learning models custom-made for this product, and none of the models we used were trained on relevant in-domain data. Instead, we composed the product from a chain of ready-made models and services.

Our decisions to do so were motivated by data wherever it was applicable, and by time-to-market considerations otherwise. Though one might argue that the VTT decision was wrong as the discarded model performance on in-domain data was better than the TPV tool we used, this is a non-issue since the end product has proved beneficial despite the shortcoming of this element in the chain of models. Moreover, making the product available to our users - which would have been blocked without these ready-made models - is a crucial element in building more accurate models in the future.

Development of the entire end-to-end process took about four months. From the time already spent on developing models for the MT, NER, and TC tasks by other teams, and the time spent on the VTT task and improving on the TC model by our team, we estimate that the development of the same product from scratch would have taken approximately two years if taken up by a single team.

Deploying the voice assistant has benefited the company business metrics two-fold, both by increasing engagement and reservations, but also by reducing the work for customer service representatives, as users found the solutions to their problems more easily when using the voice free-form interface.

To conclude our observations, we recommend to break down complex machine learning architectures into atomic sub-tasks. Contrary to an initial urge to develop a novel tailor-made solution, we found that re-purposing existing solutions can often achieve effective results while efficiently saving development and scaling efforts.
Moreover, reusable system components drive long-term system alignments and achieve services and organizational synergy.

We invite our peers to be aware of the option of building machine learning-centered products without ever having to train a single model, but rather to save valuable time and effort by using the work already done by peers and colleagues.

\begin{acks}
We would like to thank Steven Baguley, Nam Pham, Niek Tax, Guy Nadav and David Konopnicki for their feedback during writing of this paper. We also thank Chalom Asseraf, Noa Barbiro, Rubens Sonntag, Giora Shcherbakov, Dor Samet, Teri Hason and the rest of the contributors to the Voice assistant product at Booking.com.
\end{acks}

\bibliographystyle{ACM-Reference-Format}

\bibliography{sample-base}

\end{document}